\documentclass[aps,prl,twocolumn,floatfix, preprintnumbers]{revtex4}
\usepackage[pdftex]{graphicx}
\usepackage[mathscr]{eucal}
\usepackage[pdftex]{hyperref}
\usepackage{amsmath,amssymb,bm,yfonts}
\usepackage[utf8]{inputenc}

\newif\ifarxiv
\arxivfalse

\begin		{document}

\def\W {\mathcal W}

\def\section	#1{\quad\textit{#1}.---}
\def\G {h}
\def\suck[#1]#2{\includegraphics[#1]{#2}}        

\preprint{BROWN-HET-1760}

\title
    {
    Nonlinear evolution of the AdS$_4$ black hole bomb
    }

\author{Paul~M.~Chesler}
\affiliation
    {Black Hole Initiative, Harvard University, Cambridge, MA 02138, USA}
\email{pchesler@g.harvard.edu}

\author{David~A.~Lowe}
\affiliation
    {Department of Physics, Brown University, Providence, RI 02912, USA }
\email{lowe@brown.edu}

\date{\today}

\begin{abstract}
The superradiant instability of rotating black holes with negative cosmological constant is studied by 
numerically solving the full 3+1-dimensional Einstein equations. We find evidence for an epoch dominated by a solution with a single helical Killing vector and a multi-stage process with 
distinct superradiant instabilities.
\end{abstract}


\maketitle
\parskip	2pt plus 1pt minus 1pt

\section{Introduction} Energy may be extracted from rotating black holes via scattering involving superradiant modes \cite{zeldovich}. It was suggested some time ago that if such modes could be confined using a mirror, then an amplification process can occur, leading to a significant fraction of the rest mass of the black hole being converted into radiation \cite{zeldovich,Press:1972zz}.  For a review see \cite{Brito:2015oca}.
In the present work we study this process with gravitational waves and replace the mirror by a negative cosmological constant and asymtotically AdS boundary conditions.  The geometry then
contains a time-like boundary from which gravitational waves can reflect off and the resulting Kerr-AdS black holes
\cite{carter1968} can be unstable \cite{Cardoso:2004hs}. 
Aside from numerical convenience, the asymtotically AdS boundary conditions 
also mean that the system has a dual holographic interpretation as a strongly
coupled quantum field theory living on the boundary
\cite{Maldacena:1997re}.

Earlier work studied this problem perturbatively around the Kerr-AdS solution, where the quasi-normal modes have been studied in detail \cite{Cardoso:2013pza}. However up until now it has been difficult to study the nonlinear amplification mechanism in the full 3+1-dimensional case.
Consequently, the final fate of the instability remains unknown.
Progress has been made in some analog problems with more symmetry, such as Reissner-Nordstrom black holes coupled to charged scalars \cite{Sanchis-Gual:2015lje,Bosch:2016vcp} with spherical symmetry, and
massive complex vector bosons coupled to Kerr with axisymmetric geometry \cite{East:2017ovw}.

For given mass and angular momentum, it was demonstrated in Ref.~\cite{Dias:2015rxy}
that there exists ``black resonator'' solutions to Einstein's equations with higher
entropy than Kerr-AdS.  These are time-periodic black hole solutions with
a single helical Killing vector.  Since these solutions have
higher entropy, it is possible they could be the endpoint of the instability.
However, they are also unstable to superradiant instabilities \cite{Green:2015kur}.
This led Ref.~\cite{Niehoff:2015oga} to speculate the evolution of the system could
be that of a cascade, with the Kerr-AdS geometry transitioning to a black resonator,
which then subsequently transitions to another black resonator and so on.
Time evolution could lead to ever increasing structure on short distance scales, potentially violating cosmic censorship \cite{Niehoff:2015oga}.

We numerically evolve the $3+1$ dimensional Einstein equations
and find evolution consistent with the transition of Kerr-AdS to a black resonator.
Specifically, we find that the Kerr-AdS geometry undergoes the 
expected superradiant instability, but then transitions to a
solution with an approximate helical Killing vector. This solution then undergoes a new and distinct superradiant instability, 
with the most unstable mode having both shorter 
wavelength and smaller growth rate than that in Kerr-AdS. 
In the dual field theory, we find that the black resonator instability leads to an
exotic state, with regions of negative energy persisting for long periods of time.
We do not evolve to the endpoint of the black resonator instability, leaving the 
issue of the endpoint unresolved.

\section{Setup}We numerically solve the vacuum Einstein equations with negative cosmological 
constant $\Lambda = -3/L^2$ with $L$ the AdS radius, which we set to one.
Our numerical algorithm is largely described in \cite{Chesler:2013lia}.
We employ a characteristic evolution scheme with metric anzatz
\vspace{-0.15cm}
\begin{equation}
\label{eq:metric}
ds^2 = r^2 g_{\mu \nu}(x^\alpha,r) dx^\mu dx^\nu + 2 dt dr,
\end{equation}
with Greek indices $(\mu,\nu)$ running over the AdS boundary spacetime coordinates 
$x^\mu = \{t,\theta,\varphi\}$
where $\theta$ and $\varphi$ are the usual polar and azimuthal angles in spherical coordinates.  
The coordinate $r$ is the AdS radial coordinate, with the AdS boundary located at $r = \infty$.  The ansatz (\ref{eq:metric})
is invariant under the residual diffeomorphism $r \to r  + \xi(x^\alpha)$ with arbitrary $\xi(x^\alpha)$.  We exploit this  
to fix the location of the apparent horizon to be at $r = 1$ and restrict the computation domain to $r \geq 1$.

The AdS boundary is time-like and hence requires the imposition of boundary conditions there.
Near the AdS boundary one can solve Einstein's equations with a power series expansion.  Doing so 
one obtains $g_{\mu \nu}(x,r) = g_{\mu \nu}^{(0)}(x) + \dots +g_{\mu \nu}^{(3)}(x)/r^3 + \dots$.  
The expansion coefficient $g_{\mu \nu}^{(0)}$ corresponds to the metric on AdS boundary, which we choose to be that of 
the unit sphere $h_{\mu \nu} = {\rm diag}(-1,1,\sin^2 \theta)$.  
A convenient diffeomorphism invariant observable is the stress tensor in the dual boundary 
quantum field theory, which reads 
reads \cite{deHaro:2000vlm}
\begin{equation}
T_{\mu \nu} = g_{\mu \nu}^{(3)} + \textstyle \frac{1}{3}h_{\mu \nu} g_{0 0}^{(3)}.
\end{equation}

\begin{figure*}[ht!]
   \begin{center}
     \includegraphics[trim= 80 220 90 120,clip,scale=0.2]{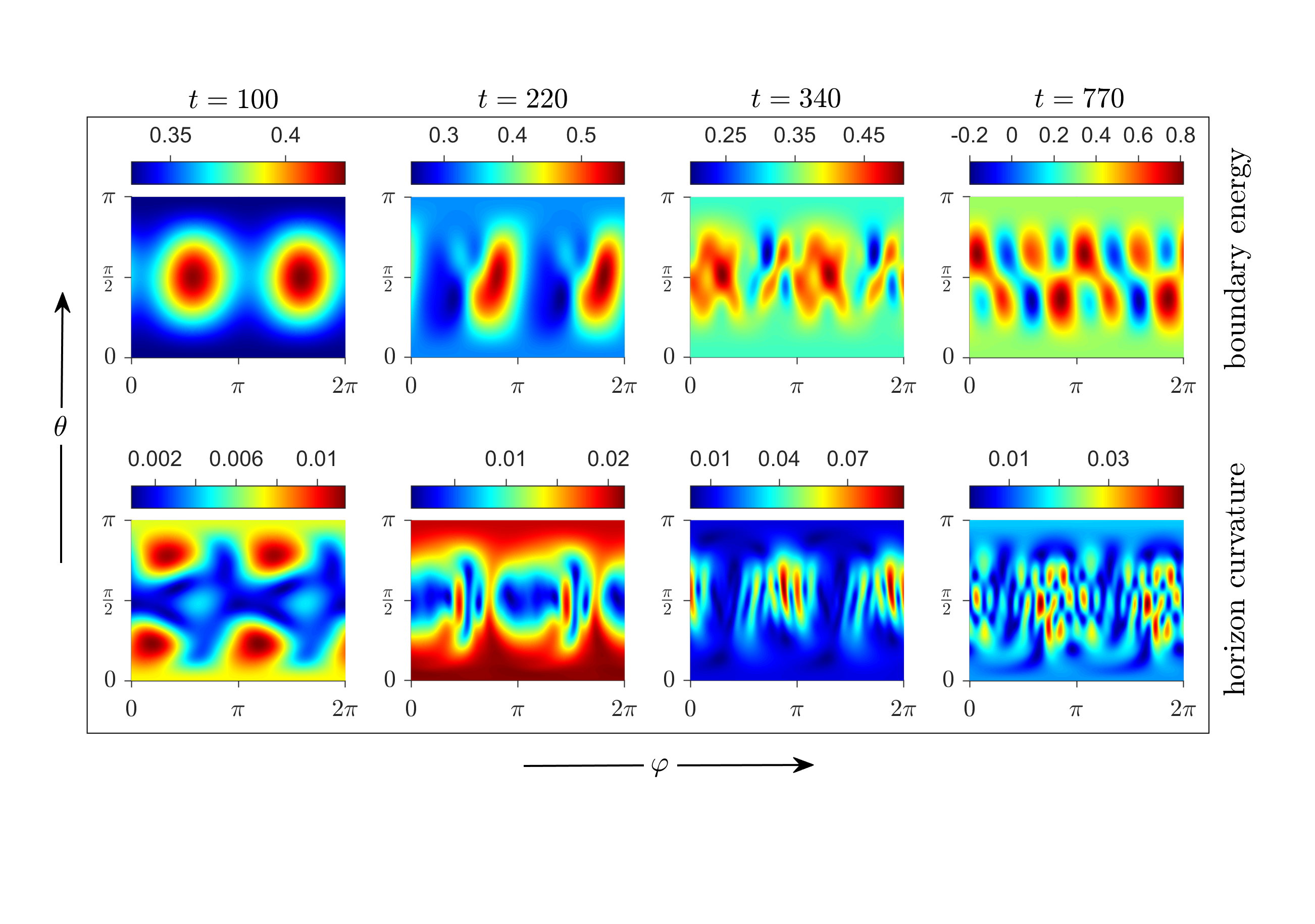}
     \caption{Top: the boundary energy density $T^{00}$ at four different times.  Bottom: 
     $\sqrt{K_{MN} K^{MN}}$, with
                 $K_{MN}$ the horizon extrinsic curvature, plotted at the same times.  
                 }
     \label{fig:EnergyAndCurvature}
   \end{center}
 \end{figure*}

With our characteristic evolution scheme, sufficient initial data consists of the normalized spatial metric 
$\hat g_{ij} \equiv g_{ij}/\sqrt{\det g_{ij}}$ with  $\{i,j\}=\{\theta,\varphi\}$,
and the boundary densities $T^{0 \mu}$
\cite{Chesler:2013lia}.  The remaining components of the metric 
are determined by constraint equations.  
For initial data we choose $\hat g_{ij}$ to be given by the Kerr-AdS metric.  We choose $T^{0 \mu}$ to be that determined by the Kerr-AdS metric, plus small perturbations to seed superradiant
instabilities.  Specifically, we add to the Kerr-AdS result for $T^{0\mu}$ small perturbations with angular momentum
$\ell = 2,3$ and azimuthal quantum number $m = 2,3$.  These perturbations do not change $\int  d \Omega \, T^{0 \mu}$, and hence the mass or angular momentum 
of the Kerr-AdS solution. We choose mass $M = 0.2783$ and angular momentum $J = 0.06350$.
We evolve 915 AdS radii in time.  
We discuss our discretization scheme in the supplemental material.

\section{Results and discussion}
In the top panel of Fig.~\ref{fig:EnergyAndCurvature} we plot snapshots of 
boundary energy density $T^{00}$ at times $t = 100, 240,340$ and $770$.
Note that the color scaling is different at each time. 
As we shall discuss further below, during $100 \lesssim t \lesssim 400$ the energy density rotates in $\varphi$ at 
angular velocity $\W \approx 1.7$.  Over longer time scales structure gradually builds up.  
At  $t = 100$ there is clearly a mode with azimuthal quantum number $m =2$ excited, with two peaks 
separated by angle $\Delta \varphi \approx \pi$.  The amplitude of this mode is initially small,
with the energy density deviating from uniformity in $\varphi$ by only $\sim 15\%$ at $t = 100$.
At $t = 220$, the two maxima have become distorted and grown in amplitude.  At $t = 340$, additional
maxima have begun to form.  By $t = 770$ there is clearly an $m = 4$
excitation in the energy density, with four maxima and minima 
both above and below the equator, with adjacent maxima separated by $\Delta \varphi \approx \pi/2$.  
In contrast to time $t = 100$, 
at $t = 770$ the variation in the energy density is large.  Indeed, the energy density in the 
minima is negative.  Negative energy density indicates an exotic state of matter in the dual
quantum field theory.  Regions of negative energy density have also been seen in the holographic 
theories studied in \cite{Athanasiou:2010pv,Casalderrey-Solana:2013aba,Chesler:2015fpa,Horowitz:2014hja}.

\begin{figure*}
   \begin{center}
     \includegraphics[trim= 10 0 10 10,clip,scale=0.60]{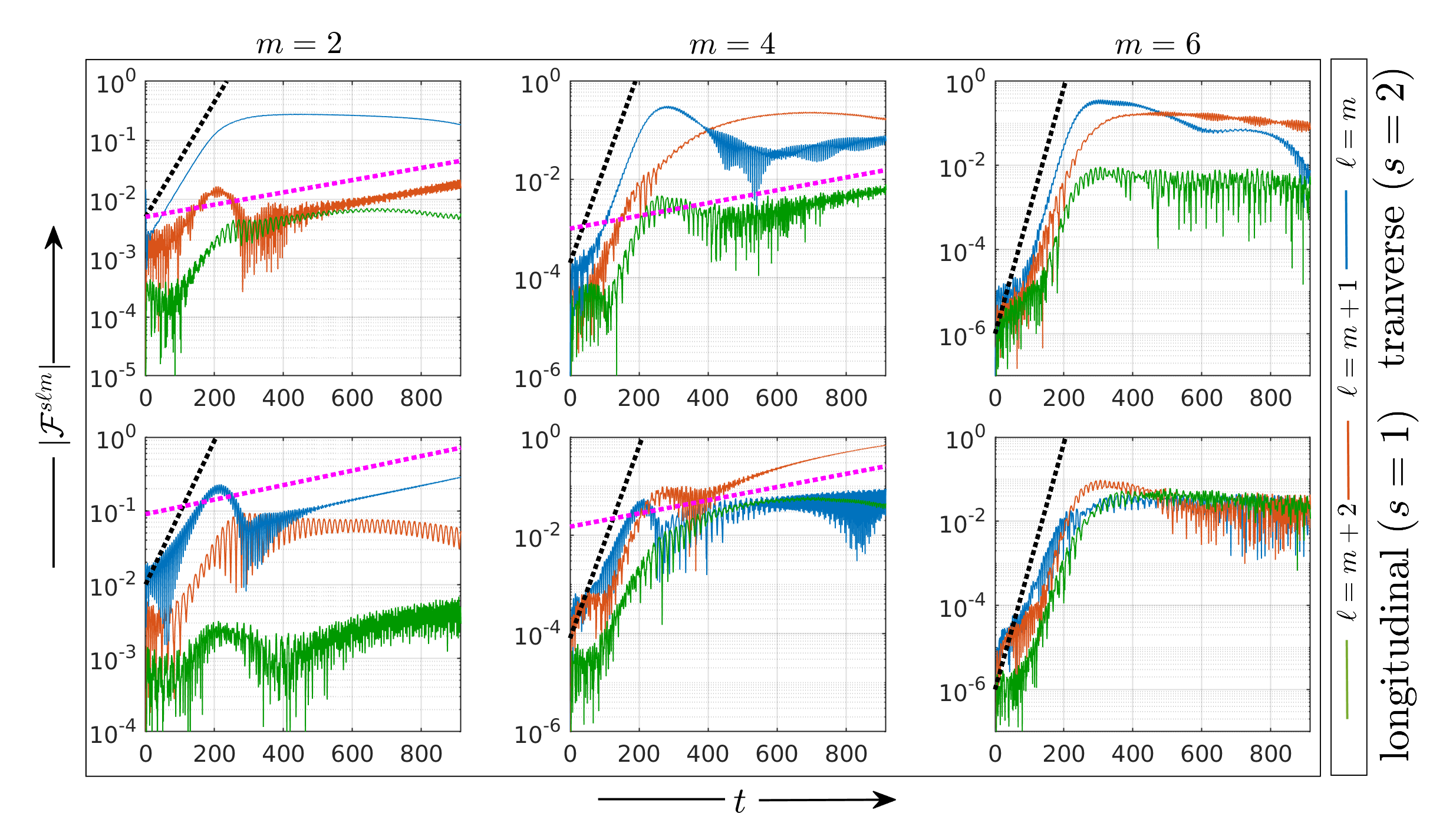}
     \caption{
                 $|\mathcal F^{s \ell m}|$ as a function of time for $m = 2,4,6$ and $\ell = m,m+1$ and $m+2$.  
                 Note the 
                 appearance of two distinct periods of exponential growth 
                 before $t \sim 200$ and after $t \sim 300$.
                 Also shown as dashed black lines
                 are $ e^{{\rm Im}\, \omega_{222} t}$,  $e^{2 {\rm Im}\, \omega_{222} t}$ 
                 and  $e^{3 {\rm Im}\, \omega_{222} t}$ for $m = 2,4,6$,          
                 respectively, with $\omega_{222}$ the complex frequency of the transverse $\ell = m = 2$ 
                 unstable mode of Kerr-AdS.
                 In the transverse channel, the pink dashed lines show $e^{0.0024 t}$ for $m = 2$ 
                 and $e^{0.003t}$ for $m = 4$, respectively.
                 In the longitudinal channel the dashed pink lines show
                 $e^{0.0023 t}$ and $e^{0.0031 t}$ for $m = 2,4$, respectively.  
}
     \label{fig:Spectrum}
   \end{center}
 \end{figure*}

The growth of structure is also evident on the horizon.  One quantity to consider is the extrinsic curvature $K_{MN}$ of the 
event horizon (with upper case latin indices running over bulk spacetime coordinates).  The extrinsic curvature  can be constructed from the null normal $n_M$ to the horizon 
and an auxiliary null vector $\ell_M$ chosen to satisfy $\ell_M n^M = -1$.  The extrinsic 
curvature is then given by $K_{MN} =  \Pi^P_{\ M} \Pi^Q_{\ N} \nabla_P n_Q$ where the projector $\Pi^{M}_{\ N} = \delta^{M}_{ \ N} + \ell^M n_N$.  As the horizon is at $r \approx 1$ 
\footnote{Indeed with this choice of $n_M$ we find $n_M n^M = O(10^{-4})$.},
we choose $n_M dx^M = dr$ and $\ell_M dx^M = -dt$.  In the bottom panel of Fig.~\ref{fig:EnergyAndCurvature} we plot snapshots of $\sqrt{K^{MN}K_{MN}}$, again at times $t = 100, 240,340$ and $770$.  

 To elucidate the growth of different modes we study the evolution of the  
 spherical harmonic transform of the boundary momentum density 
  \begin{align}
 \mathcal{F}^{s\ell m} \equiv \int d\Omega \, \mathcal V^{*s \ell m}_i T^{0i},
 \end{align}
 with vector spherical harmonics $\mathcal V^{s \ell m}_i$ given by 
 \begin{align}
 &\mathcal V_i^{1\ell m} = {\textstyle \frac{1}{\sqrt{\ell (\ell + 1)}}} \nabla_i y^{\ell m}, &
\mathcal V_i^{2\ell m} =  {\textstyle \frac{1}{\sqrt{\ell (\ell + 1)}}} \epsilon_i^{\ j} \nabla_j y^{\ell m}.&
\end{align}
Here 
$\epsilon_i^{\ j}$ has non-zero components $\epsilon_\theta^{\ \varphi} = \csc \theta$ and $\epsilon_{\varphi}^{\ \theta} = -\sin \theta$, and $y^{\ell m}$ are spherical harmonics.
The $s = 1,2$ modes encodes the longitudinal and
transverse component of the momentum density, 
respectively.

\begin{table}[t!]
\begin{tabular}{@{\extracolsep{5pt}}c|cccccc}
{\rm channel} & $\mathcal F^{122}$ & $\mathcal F^{222}$ & $\mathcal F^{133}$ & $\mathcal F^{233}$ & $\mathcal F^{144}$ & $\mathcal F^{244}$ 
\\
\hline
\hline
${\rm Re} \, \omega $ & 2.52  & 3.38 & 3.74 & 4.55 & 4.83 & 5.64
\\[2pt]
${\rm Im} \, \omega \times 10^2 $ & $0.424$ & $2.25$ & $0.0218$ & $0.136$ & $0.00108$ & $0.00710$ 
\end{tabular}
\caption
{%
Complex frequencies for some unstable modes in Kerr-AdS with mass $M = 0.2783$ and angular momentum $J = 0.06350$.
}
\label{T1}
\end{table}

In Fig.~\ref{fig:Spectrum} we plot $|\mathcal F^{s \ell m}|$ for $m = 2,4,6$ (left, middle, right columns)
and $\ell = m$ (blue), $\ell = m+1$ (red) and $\ell = m+2$ (green).  
The top row of plots shows the transverse channel $|\mathcal F^{2 \ell m}|$ while the
bottom row of plots shows the longitudinal channel $|\mathcal F^{1 \ell m}|$.  
The rapid oscillations seen in the plots are not numerical noise, but rather 
due to structure present on short time scales.
Over large time scales there are two distinct periods of growth in $\mathcal F^{s \ell m}$,
which we shall refer to as primary and secondary.  The primary growth 
occurs until $t \sim 200$ and the secondary growth occurs after $t \sim 300$
and is slower than the primary growth.  As we shall see below,
the primary growth is driven by the leading linear instability in the 
Kerr-AdS geometry, the $\ell = m = 2$ mode in the transverse channel.

A linearized mode analysis about our choice of Kerr-AdS black hole 
yields a tower of unstable modes parameterized by $\ell$ and $m$
in both the transverse and longitudinal channels \cite{Cardoso:2013pza}.  The associated 
complex frequencies for the most unstable modes are given in Table~\ref{T1}.  From the table we see 
that the dominant unstable mode is that of the $\ell = m = 2$ mode in the transverse channel with complex frequency $\omega_{222} \approx 3.38 + 0.025 i$.
This mode grows more than 5 times faster than the next subdominant mode.
Also included in Fig.~\ref{fig:Spectrum} are plots of $e^{{\rm Im}\, \omega_{222}\, t}$, $e^{2 {\rm Im}\, \omega_{222}\, t}$ and $e^{3{\rm Im}\, \omega_{222}\, t}$ in the $m=2,4,6$ columns, 
respectively.  
These curves are shown as the dashed black lines in each plot.
During the primary growth we clearly see 
$|\mathcal F^{222}| \sim e^{{\rm Im}\, \omega_{222}\, t}$, indicating growth consistent with a linearized analysis.
At $t \sim 200$, when the primary growth begins to slow, $|\mathcal F^{222}|$ has grown to $\sim 0.1$.
Evidently, the linearized analysis breaks down at this 
point, with $|\mathcal F^{222}|$ subsequently plateauing and slowly decreasing thereafter.
For the $m=4$ and $m = 6$ modes, we see $|\mathcal F^{244}| \sim e^{2{\rm Im}\, \omega_{222}\, t}$ and  
$|\mathcal F^{266}| \sim e^{3{\rm Im} \, \omega_{222}\, t}$ until $t \sim 200$, after which
both modes begin to decrease in magnitude.  This behavior is 
expected from perturbation theory, 
where quadratic and cubic nonlinearities of the dominant $m=2$ mode source $m = 4$ and $m = 6$ modes
with amplitudes scaling like the square and cube of $ e^{{\rm Im} \,\omega_{222}\, t}$, respectively. 

The $\ell = m$ modes in the longitudinal channel
also grow until $t \sim 200$, albeit with large and rapid oscillations in the amplitudes.
This suggests that the primary stage growth of longitudinal modes is also driven by nonlinear interactions with transverse modes.
Indeed, the amplitude of the $m=2,4$ and $6$ longitudinal modes
roughly scales like $ e^{{\rm Im} \,\omega_{222}\, t}$, $ e^{2{\rm Im} \,\omega_{222}\, t}$ and $ e^{3{\rm Im} \,\omega_{222}\, t}$, respectively. 

We note that the presence of a growing $m =2$ longitudinal 
mode is consistent with our plots of the energy density in Fig.~\ref{fig:EnergyAndCurvature}, 
where at $t = 100$ and $220$, the dominant excitation was $m = 2$.  
We also note that at all times studied in this Letter,
we see no significant growth in modes with odd values of $m$.
This is consistent with the small growth rates for $m = 3$ appearing 
in Table~\ref{T1}, and with the fact that in perturbation theory
an $m =2$ mode does not source modes with $m$ odd.

We now turn to the secondary growth seen in Fig.~\ref{fig:Spectrum}.
Also shown in Fig.~\ref{fig:Spectrum} as dashed pink lines in the transverse channel 
are plots of $e^{0.0024 t}$ and $e^{0.003t}$ for $m = 2,4$, respectively.
In the longitudinal channel we include plots 
of $e^{0.0023 t}$ and $e^{0.0031 t}$ for $m = 2,4$, respectively,
again shown as dashed pink lines.
Evidently, after $t \sim 300$ we 
have $|\mathcal F^{232}| \sim e^{0.0024 t}$ and $|\mathcal F^{122}| \sim e^{0.0023 t}$,
with $|\mathcal F^{122}| \sim 0.3$ at $t = 915$.  
The $\ell = 5$, $m = 4$ longitudinal mode shows the strongest secondary growth, albeit not
purely exponential, with $| \mathcal F^{154}| \approx 0.7$ and the approximate
tangent $e^{0.0031 t}$ at $t = 915$.
A large $m = 4$ secondary mode is consistent with our plots of the energy density  
in Fig.~\ref{fig:EnergyAndCurvature}, where by 
$t = 770$ we saw an order one amplitude $m = 4$ perturbation in the energy density.

It should be emphasized that the growth rates associated
with the secondary instability do not coincide with any of the complex frequencies 
given in Table~\ref{T1}.  This is to be expected since by the time 
secondary instabilities kick in, the geometry is very different from Kerr-AdS,
from which Table~\ref{T1} was obtained.

 \begin{figure}[h]
     \includegraphics[trim= 10 20 30 40,clip,scale=0.16]{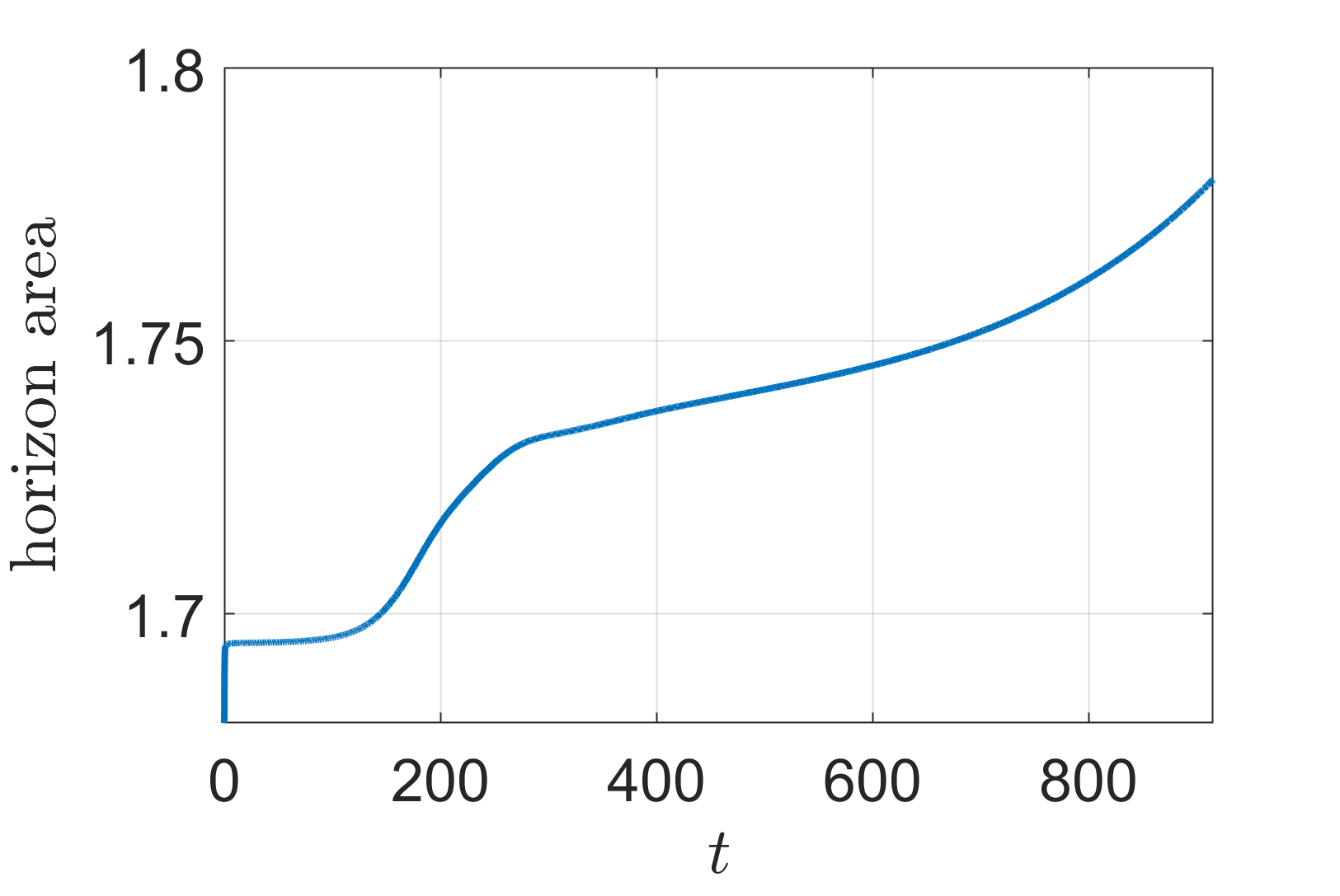}
     \caption{The horizon area as a function of time.  After initial transients decay, the horizon settles
     down to its Kerr-AdS value $1.694$.  It subsequently undergoes
     two distinct periods of growth, with the second period ongoing at $t = 915$.}
     \label{fig:HorizonArea}
 \end{figure}
 
The distinct primary and secondary growth rates also manifest themselves 
in the horizon area, which we plot in Fig.~\ref{fig:HorizonArea}.  After initial transients decay,
the horizon area is $\approx 1.694$, which is the area of our initial Kerr-AdS black hole.  The horizon 
area grows rapidly around  $t \sim 200$ and dramatically slows down after $t \sim 300$. 
However, by  $t \sim 600$, the horizon area growth accelerates again.

What is the physics which leads to two distinct epochs with different instability growth rates?
As mentioned above, for given mass and angular momentum
there exist hairy black hole solutions with larger entropy than Kerr-AdS, but with only a 
single \textit{helical} Killing vector 
\cite{Dias:2015rxy}
\begin{equation}
\label{eq:Killing}
K = \partial_t + \W \partial_{\varphi},
\end{equation}
with $\W$ a constant.  Such solutions
should naturally be generated by superradiant 
instabilities when there is a gap in the 
growth rates, with a single dominant unstable mode
\cite{Dias:2011at}. Indeed, for a single mode
$e^{-i \omega t + m \varphi}$, Eq.~(\ref{eq:Killing})
is an approximate Killing vector with 
\begin{equation}
	\label{eq:Killing2}
	\W = {\rm Re}(\omega)/m. 
\end{equation}
From Table~\ref{T1} we see that the dominant unstable mode has
${\rm Re}(\omega_{222})/2 \approx 1.7.$
Black resonators with $\W > 1$ are themselves unstable to superradiant instabilities \cite{Green:2015kur}.
Hence, if the primary superradiant instability leads to a black resonator with $\W \approx 1.7$, there should also be a subsequent secondary superradiant instability.

\begin{figure}[h]
     \includegraphics[trim= 0 0 90 50,clip,scale=0.15]{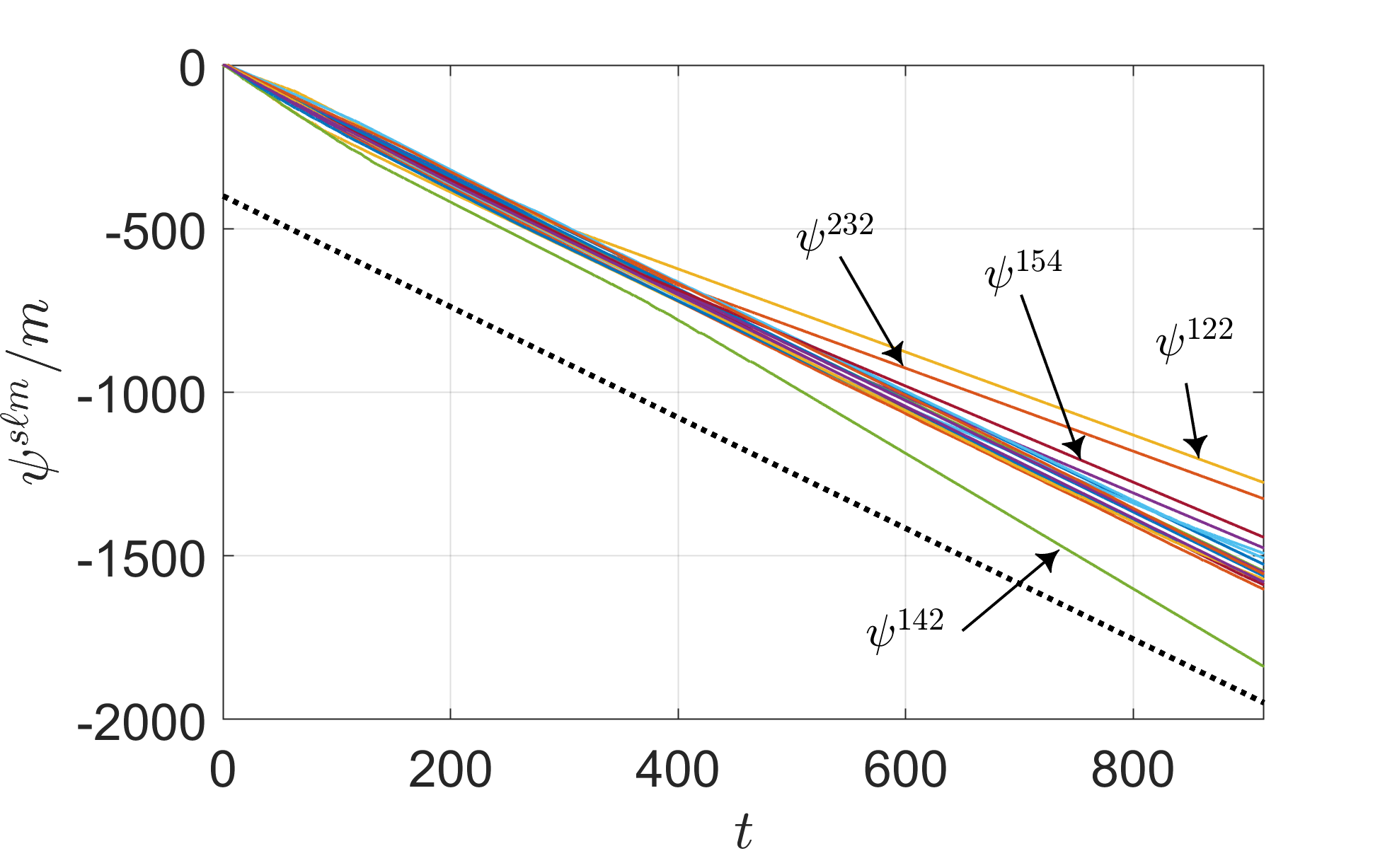}
     \caption{Phase angles normalized by $m$ for the modes plotted in Fig.~\ref{fig:Spectrum}.
                 All curves have the same slope during $ 100 \lesssim t \lesssim 400$.
                 The dashed line, shown for reference, has slope $-1.7$.
                 }
     \label{fig:phase}
\end{figure}

To explore whether our numerics are consistent with a transition to
a black resonator, 
we look for an approximate helical Killing vector.  If (\ref{eq:Killing}) is a Killing vector,
then the metric and boundary stress tensor should 
only depend on the combination $\varphi - \W t$, meaning the entire solution
rotates at constant angular velocity. 
Writing $\mathcal F^{s\ell m} = |\mathcal F^{s\ell m}| e^{i \psi^{s\ell m}}$, we therefore expect 
phases $ \psi^{s\ell m} = - m \W t + \rm const.$  In Fig.~\ref{fig:phase}
we plot $\psi^{s\ell m}/m$ for the same modes shown in Fig.~\ref{fig:Spectrum}.  
The black dashed line shows the Kerr-AdS 
prediction $\W = 1.7$.
For $ 100 \lesssim t \lesssim 400$, 
we see that all curves have the same slope $\approx -1.7$. 
Similar results can be obtained by studying the bulk geometry.
Evidently, the dominant transverse $\ell = m = 2$ mode 
in the Kerr-AdS geometry drives the system to a black resonator configuration with an approximate Killing vector determined by 
(\ref{eq:Killing}) and (\ref{eq:Killing2}). 
Given that the associated black resonator
has its own superradiant instability, it is natural
that we see a secondary instability develop with different growth rates.

What then is the final fate of the system? Does the 
secondary instability drive the system to another black
resonator, with a subsequent black resonator cascade developing as speculated in \cite{Niehoff:2015oga}?
To have this happen one presumably needs 
a gap in the secondary growth rates, so that a single 
dominant mode $e^{-i \omega t + m \varphi}$
will grow and the solution will have 
(\ref{eq:Killing}) as an approximate
Killing vector with $\W$ determined by (\ref{eq:Killing2}). In contrast to the Kerr-AdS 
superradiant instability, where the dominant unstable
mode grew approximately five times faster than the next
subdominant mode,
the secondary instabilities we observe don't have a large gap in the growth rates.  Indeed, as discussed above, at $t = 915$ we have $|\mathcal F^{122}| \sim e^{0.0023 t}$
and $|\mathcal F^{154}| \sim e^{0.0031 t}$. From
Fig.~\ref{fig:phase} we also see that after $t \sim 400$
we have $\psi^{122}/2 \sim -1.27 t$
and $\psi^{154}/4 \sim -1.47 t$, indicating 
that these two modes rotate at angular frequencies 
differing by $\sim 15\%$. This suggests that the 
secondary instability doesn't lead to another 
black resonator, but instead to a hairy black
hole with no symmetries.  

To ascertain the final fate of the system, 
longer duration simulations must be performed.
It should be possible to evolve further
using our characteristic evolution scheme, as we see no
fundamental problems with the numerics, such as the development 
of caustics.  However, doing so will require lengthy run times because the superradiant growth 
rates are very small, and as more structure develops, higher resolutions are needed.
We leave this for future work.

\section{Acknowledgments}%
PC is supported by the Black Hole Initiative at Harvard University, 
which is funded by a grant from the John Templeton Foundation. 
DL is supported in part by DOE grant DE-SC0010010.
We thank Jorge Santos for useful conversations and for providing
the data shown in Table~\ref{T1}.

\bibliographystyle{utphys}
\bibliography{refs}%
%
\pagebreak
\widetext
\begin{center}
\textbf{\large Supplemental Materials: Nonlinear evolution of the AdS$_4$ black hole bomb}
\end{center}

\vspace{1cm}

\setcounter{equation}{0}
\setcounter{figure}{0}
\setcounter{table}{0}
\setcounter{page}{1}

Following \cite{Chesler:2013lia}, we employ  an inverse radial coordinate $u \equiv \frac{1}{r} \in (0,1)$
and expand the $u$ dependence of all functions in a pseudo-spectral basis 
of Chebyshev polynomials.
We employ domain decomposition 
in the $u$ direction with four domains containing $N_u = \{6,25,25,25\}$ points (with the least populated domain lying next to the AdS boundary).  The domain interfaces lie at $u = \{0.1,0.4,0.7\}$.

For the $(\theta,\varphi)$ dependence we employ a basis of 
scalar, vector and tensor harmonics.
These are eigenfunctions of the covariant Laplacian $-\nabla^2$ on the unit sphere.
The scalar eigenfunctions are just 
spherical harmonics $y^{\ell m}$.  There are two vector harmonics, $\mathcal V_i^{s \ell m}$ with $s = 1,2$,
and three symmetric tensor harmonics, $\mathcal T_{ij}^{s \ell m}$, $s = 1,2,3$.  Explicit representations 
of these functions are easily found and read \cite{oldref}
\begin{subequations}
\begin{eqnarray}
 \mathcal V_i^{1\ell m} &=& {\textstyle \frac{1}{\sqrt{\ell (\ell + 1)}}} \nabla_i y^{\ell m}, \\
\mathcal V_i^{2\ell m} &=&  {\textstyle \frac{1}{\sqrt{\ell (\ell + 1)}}} \epsilon_i^{\ j} \nabla_j y^{\ell m},\\
\mathcal  T_{ij}^{1 \ell m} &=& {\textstyle \frac{\G_{ij}}{\sqrt{2}}} y^{\ell m}, 
 \\ 
 \mathcal T_{ij}^{2\ell m} &=& {\textstyle \frac{1}{ \sqrt{\ell (\ell + 1)( \ell(\ell + 1)/2 - 1)}}} 
  \epsilon_{(i}^{\ \  k} \nabla_{j)} \nabla_k   y^{\ell m},
  \\
 \mathcal T^{3 \ell m}_{ij} &=& {\textstyle \frac{1}{\sqrt{\ell (\ell + 1)( \ell(\ell + 1)/2 - 1)}}} [ \nabla_i \nabla_j 
 + {\textstyle\frac{\ell (\ell + 1)}{2}} \G_{ij} ]  y^{\ell m}, \ \ \ \  \ \ \ 
\end{eqnarray}
\end{subequations}
where $\epsilon_i^{\ j}$ has non-zero components $\epsilon_\theta^{\ \varphi} = \csc \theta$ and $\epsilon_{\varphi}^{\ \theta} = -\sin \theta$, and $h_{ij} = {\rm diag}(1,\sin^2 \theta)$ is the metric 
on the unit sphere.
The scalar, vector and tensor harmonics are orthonormal and complete.

We expand the metric as follows,
\begin{subequations}
\label{eq:expansions}
\begin{eqnarray}
g_{00}(t,u,\theta,\varphi)  &=& \sum_{\ell m}  \alpha_{\ell m}(t,u) y^{\ell m}(\theta,\varphi),
\\
g_{0i}(t,u,\theta,\varphi)  &=& \sum_{s\ell m}  \beta^{s\ell m}(t,u) \mathcal V^{s\ell m}_i(\theta,\varphi),
\\
g_{ij}(t,u,\theta,\varphi)  &=& \sum_{s\ell m}  \gamma^{s\ell m}(t,u) \mathcal  T^{s\ell m}_{ij}(\theta,\varphi).
\end{eqnarray}
\end{subequations}
A analogous expansion can be written for the boundary stress tensor $T^{\mu \nu}$.
Derivatives in $\{\theta,\varphi\}$ can then be taken by differentiating the scalar, vector and tensor harmonics.

In order to efficiently transform between real space and mode space, we employ a Gauss-Legendre grid in $\theta$ with $\ell_{\rm max} + 1$ points.  Likewise, we employ a Fourier grid in the $\varphi$ direction with $2 \ell_{\rm max} + 1$ points.
These choices allow the transformation between mode space and real space to be done with a combination 
of Gaussian quadrature and Fast Fourier Transforms.
 
We truncate the expansions (\ref{eq:expansions}) at maximum angular momentum $\ell_{\rm max} = 39$.  To test the convergence of our numerics, we have also ran identical simulations with $\ell_{\rm max} = 30$ and obtained plots indistinguishable from those presented in this Letter.  To illustrate this, below we plot the mode amplitudes 
$|\mathcal F^{s\ell m}|$ shown in Fig.~2 in the Letter.  The dashed black lines show the same curves obtained with $\ell_{\rm max} = 30$ instead of $\ell_{\rm max} = 39$.  In order show fine structure, we have restricted the 
time interval to $t \in (200,500)$.  As is evident from the Figure, the $\ell_{\rm max} = 39$ and $\ell_{\rm max}= 30$ data sets agree very well, with the two sets of curve indistinguishable from each other.   This behavior continues throughout our entire computation time.

\begin{figure*}[h]
   \begin{center}
     \includegraphics[trim= 30 0 10 10,clip,scale=0.25]{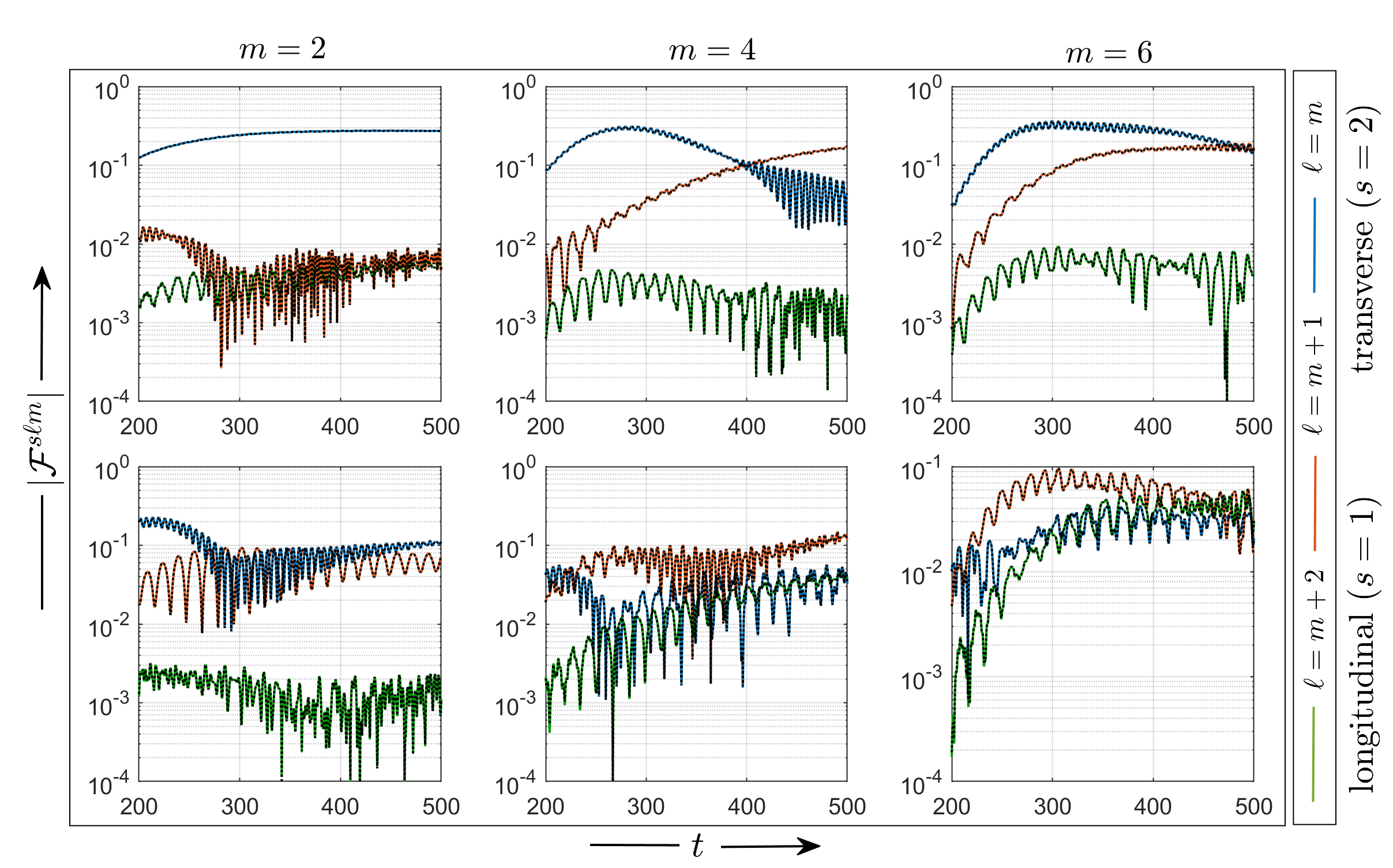}
     \caption{
                 $|\mathcal F^{s \ell m}|$ as a function of time for $m = 2,4,6$ and $\ell = m,m+1$ and $m+2$.  
                 The solid lines show results obtained with $\ell_{\rm max} = 39$ while the dashed black curves show
                 results obtained with $\ell_{\rm max} = 30$.
}
     \label{fig:convergence}
   \end{center}
 \end{figure*}

\end{document}